\documentclass[twocolumn]{aastex6}
\usepackage{natbib}

\begin{document}

\title{A Targeted Search for point sources of E\MakeLowercase{e}V photons with the\\ Pierre Auger Observatory}

% created on 2016-12-02
\author{
A.~Aab\altaffilmark{63},
P.~Abreu\altaffilmark{70},
M.~Aglietta\altaffilmark{48,47},
I.~Al Samarai\altaffilmark{29},
I.F.M.~Albuquerque\altaffilmark{16},
I.~Allekotte\altaffilmark{1},
A.~Almela\altaffilmark{8,11},
J.~Alvarez Castillo\altaffilmark{62},
J.~Alvarez-Mu\~niz\altaffilmark{79},
G.A.~Anastasi\altaffilmark{38}
% ...
}

\fullcollaborationName{The Pierre Auger Collaboration}

\altaffiltext{1}{Centro At\'omico Bariloche and Instituto Balseiro (CNEA-UNCuyo-CONICET), Argentina}
% ...
\altaffiltext{8}{Instituto de Tecnolog\'\i{}as en Detecci\'on y Astropart\'\i{}culas (CNEA, CONICET, UNSAM), Centro At\'omico Constituyentes, Comisi\'on Nacional de Energ\'\i{}a At\'omica, Argentina}
% ...
\altaffiltext{11}{Universidad Tecnol\'ogica Nacional -- Facultad Regional Buenos Aires, Argentina}
% ...
\altaffiltext{16}{Universidade de S\~ao Paulo, Inst.\ de F\'\i{}sica, S\~ao Paulo, Brazil}
% ...
\altaffiltext{29}{Laboratoire de Physique Nucl\'eaire et de Hautes Energies (LPNHE), Universit\'es Paris 6 et Paris 7, CNRS-IN2P3, France}
% ...
\altaffiltext{38}{Gran Sasso Science Institute (INFN), L'Aquila, Italy}
% ...
\altaffiltext{47}{INFN, Sezione di Torino, Italy}
\altaffiltext{48}{Osservatorio Astrofisico di Torino (INAF), Torino, Italy}
% ...
\altaffiltext{62}{Universidad Nacional Aut\'onoma de M\'exico, M\'exico}
\altaffiltext{63}{Institute for Mathematics, Astrophysics and Particle Physics (IMAPP), Radboud Universiteit, Nijmegen, Netherlands}
% ...
\altaffiltext{70}{Laborat\'orio de Instrumenta\c{c}\~ao e F\'\i{}sica Experimental de Part\'\i{}culas -- LIP and Instituto Superior T\'ecnico -- IST, Universidade de Lisboa -- UL, Portugal}
% ...
\altaffiltext{79}{Universidad de Santiago de Compostela, Spain}
% ...

\begin{abstract}

Simultaneous measurements of air showers with the fluorescence and surface detectors of the Pierre Auger Observatory allow a sensitive search for EeV photon point sources. Several Galactic and extragalactic candidate objects are grouped in classes to reduce the statistical penalty of many trials from that of a blind search and are analyzed for a significant excess above the background expectation. The presented search does not find any evidence for photon emission at candidate sources, and combined $p$-values for every class are reported. Particle and energy flux upper limits are given for selected candidate sources. These limits significantly constrain predictions of EeV proton emission models from non-transient Galactic and nearby extragalactic sources, as illustrated for the particular case of the Galactic center region.

\end{abstract}

\keywords{astroparticle physics --- cosmic rays --- methods: data analysis }

\section{Introduction} 
\label{sec:intro}

Ultra-high energy (UHE) photons with energies around 1~EeV (1~EeV = $10^{18}$~eV) and above have not yet been identified (see \cite{BleveICRC} and references therein). At these high energies photons are produced primarily by $\pi^0$ decays, implying the existence of hadrons (that cause the production of $\pi^0$ mesons) with energies typically 10 times higher than the secondary photon \citep{Risse:2007sd}. At energies of about 5~EeV, around the ``ankle'' of the energy spectrum~\citep{Spectrum, ICRC2013_S}, several experiments, including the Pierre Auger Observatory, HiRes, and Telescope Array, have all found their measurements to be consistent with the existence of a light component among the cosmic rays~\citep{xsection, Xmax, PhysRevD.90.122006, Aab2016288, HiRes, TA}. If these protons were to interact in the vicinity of their sources they can produce photons by pion photoproduction or inelastic nuclear collisions. Since photons are not deflected by magnetic fields, the experimental signature would then be an accumulation of photon-like events from a particular celestial direction. 

Assuming that the energy spectra of measured TeV $\gamma$ sources \citep{Hinton:2009zz,HESSCite} extend to EeV energies, it is plausible that photon and neutron fluxes are also detectable in the EeV energy range. Sources producing particle fluxes according to an $E^{-2}$ energy spectrum inject equal energy into each decade. A measured energy flux of 1~eV~cm$^{-2}$~s$^{-1}$ in the TeV decade would result in the same energy flux in the EeV decade if the spectrum continues to such high energies and energy losses en route to Earth are negligible (see Section \ref{sec:horizon}). Furthermore, the H.E.S.S.\ collaboration measured a TeV gamma ray spectrum from the Galactic center region without any observation of a cutoff or a spectral break up to tens of TeV, implying that our Galaxy hosts petaelectronvolt accelerators called ``PeVatrons'' \citep{HESSPevatron}. If these photons are produced in hadronic processes they are necessarily accompanied by neutrons produced in charge exchange interactions of protons. The ratio between photon and neutron emissivities from $p$-$p$ collisions at the same pivot energy depends primarily on the spectral index of the proton source and it is shown in \cite{Crocker0004-637X-622-2-892} that for spectral indices $\Gamma_p \lesssim 2.4$ photon emissivities dominate, assuming a continuation of the parent proton spectrum well beyond the pivot energy. Several experiments, including the Pierre Auger Observatory, Telescope Array, IceCube, and KASCADE, searched for an indication of neutron fluxes above the PeV energy range from specific source directions, but no significant excess or correlation with catalogs could be found (e.g.\ \cite{AugerTargetedNeutron,TANeutron,KASKADENeutron,IceCubeNeutron}). 

This paper reports on a targeted search for photon point sources at EeV energies and complements previous neutron searches. The search for a photon flux, as opposed to a neutron flux, has a more direct connection to TeV measurements where the messengers are photons. A neutron flux is limited by decay of the neutrons with a mean path length of $9.2 \times E~[{\rm EeV}]$~kpc, requiring an energy of at least 1~EeV to observe the Galactic center region. In this paper we apply a lower energy threshold of $10^{17.3}$~eV using events measured by the air fluorescence detector (FD) as well as the surface detector (SD) of the Pierre Auger Observatory (see Section \ref{sec:Dataset}). This choice provides high event statistics despite the reduced duty cycle of the FD. The sensitivity to photon point sources is increased by reducing the hadronic background contribution using mass-sensitive observables. In the case of neutron-induced air showers that is not possible, since they are indistinguishable from proton primaries.

In a previous paper the Pierre Auger Collaboration published the directional search for EeV photon point sources from any direction in the exposed sky (blind search; \cite{BlindPhoton}). That analysis did not find a statistical evidence for any photon flux. The detected small $p$-values are within the expectation given 526,200 target centers. To reduce the statistical penalty of many trials from that of a blind search, this analysis focuses on just 12 target sets, each set being a class of possible sources of high-energy photons (Section \ref{sec:horizon}). The candidate sources all lie in the declination range $-85^{\circ}$ and $+ 20^{\circ}$. Targets in each class are combined in a ``stacked analysis,'' assuming that most or all candidate sources in a target set are emitting photons resulting in a more significant combined signal compared to that of a single target (Section \ref{sec:Analysis}). The results of this analysis, including particle and energy flux upper limits of selected target directions, are given in Section \ref{sec:Results}.
This study uses the same methods for hadron reduction and calculation of upper limits that were explained in the preceding paper.

\section{Data Set} 
\label{sec:Dataset}
Air showers induced by UHE cosmic rays detected with the Pierre Auger Observatory \citep{ThePierreAuger:2015rma} are used in this analysis. The observatory is located in Argentina near Malarg\"ue and is centered at latitude $35.2^\circ$~S and longitude $69.5^\circ$~W at a mean altitude of 1400~m above sea level. A SD of 1660 water-Cherenkov particle detectors covering an area of 3000~km$^2$ on a triangular grid with 1.5~km spacing observes electrons, muons, and photons at the ground with a duty cycle of nearly 100\%. The area is overlooked by 27 fluorescence telescopes operating on dark nights with a duty cycle of $\sim$15\%. Events recorded between 2005 January and 2013 December in hybrid mode, i.e., recorded by both the FD and SD, are used in this analysis. The selection criteria are the same as in the previous blind search paper \citep{BlindPhoton}, but additional accumulated data increase the statistics by 28\% to 308,676 recorded events in the present study. The energy range is between 10$^{17.3}$~eV and 10$^{18.5}$~eV to take advantage of high statistics at low energies and to avoid additional shower development processes at the highest energies \citep{Risse:2007sd}. The average angular resolution of the final dataset is 0.7$^\circ$.

\section{Target Set} 
\label{sec:horizon}
\begin{table*}[t]
  \tiny
  \begin{center}
    \caption{Combined Unweighted probabilities $\mathcal{P}$ and Weighted Probabilities $\mathcal{P}_w$ for the 12 Target Sets. \\ \textbf{Note.} In addition, information on the most significant target from each target set is given. The number of observed (Obs) and expected (Exp) events and the corresponding exposure are shown. The numbers in brackets in the observed number of events column indicate the number of events needed for a $3\sigma$ observation unpenalized and penalized ($^*$). Upper limits (UL) are computed at 95\% confidence level. The last two columns indicate the $p$-value unpenalized ($p$) and penalized ($p^*$). Due to the discrete distribution of $p$-values arising in isotropic simulations, $\mathcal{P}$ can differ from $p$ in the sets that contain only a single target.}
    \begin{tabular}{    l        l    | l l   | |  l l l l l l l l l l  l   l   }
    \hline \hline
     Class & No. & $\mathcal{P}_w$ & $\mathcal{P}$ & R.A.  & Decl. & Obs & Exp & Exposure & Flux UL & $E$-flux UL & $p$ & $p^*$  \\ 
      & & & &{\scriptsize [$^\circ$]} & {\scriptsize [$^\circ$] } &  &  & {\scriptsize [km$^{2}$~yr]} & {\scriptsize [km$^{-2}$~yr$^{-1}$] }& {\scriptsize[eV~cm$^{-2}$~s$^{-1}$] }&  &  \\  \hline
     msec PSRs & 67  & 0.57 &  0.14 & 286.4 & 4.0 & 5 (7,9$^*$) & 1.433 & 236.1 & 0.043 & 0.077 & 0.010 &  0.476  \\      
     $\gamma$-ray PSRs & 75  & 0.97 & 0.98 & 312.8 & -8.5 & 6 (8,10$^*$) & 1.857 & 248.1 & 0.045 & 0.080 & 0.007 & 0.431 \\      
     LMXB & 87 & 0.13 & 0.74 & 258.1&  -40.8 & 6 (8,11$^*$) & 2.144 & 233.9 & 0.046 & 0.083 & 0.014 &   0.718   \\      
     HMXB & 48 & 0.33 & 0.84 & 285.9 & -3.2 & 4 (7,9$^*$) & 1.460 & 235.2 & 0.036 & 0.066 & 0.040 &   0.856    \\     
     H.E.S.S. PWN & 17 & 0.92 & 0.90 & 266.8& -28.2 & 4 (8,10$^*$) & 2.045 & 211.4 & 0.038 & 0.068 & 0.104 & 0.845      \\          
     H.E.S.S. other & 16 & 0.12 & 0.52 & 258.3 & -39.8 & 5 (8,10$^*$) & 2.103 & 233.3 & 0.040 & 0.072 & 0.042 &  0.493    \\          
     H.E.S.S. UNID & 20 & 0.79 &  0.45 & 257.1 & -41.1 & 6 (8,10$^*$) & 2.142 & 239.2 & 0.045 & 0.081 & 0.014 & 0.251   \\      
     Microquasars & 13 & 0.29  &  0.48 & 267.0 & -28.1  & 5 (8,10$^*$) & 2.044 & 211.4 & 0.045 & 0.080 & 0.037 &   0.391   \\          
     Magnetars & 16 & 0.30 & 0.89 & 257.2 &  -40.1 & 4 (8,10$^*$) & 2.122 & 253.8 & 0.031 & 0.056 & 0.115 & 0.858     \\      
     Gal.\ Center & 1 &  0.59 & 0.59 & 266.4 & -29.0  & 2 (8,8$^*$) & 2.048 & 218.9 & 0.024 & 0.044 & 0.471 & 0.471   \\     
     LMC & 3 & 0.52 & 0.62 & 84.4 & -69.2  & 2 (8,9$^*$) & 2.015 & 180.3 & 0.030 & 0.053 & 0.463 & 0.845    \\       
     Cen A & 1 & 0.31 & 0.31 & 201.4 & -43.0  & 3 (8,8$^*$) & 1.948 & 214.1 & 0.031 & 0.056 & 0.221 & 0.221  \\ \hline \hline
     \end{tabular}
  \end{center}
  \label{tab:Targets}
\end{table*}

\begin{figure}[t!]
\centering
\includegraphics[width=8.6cm]{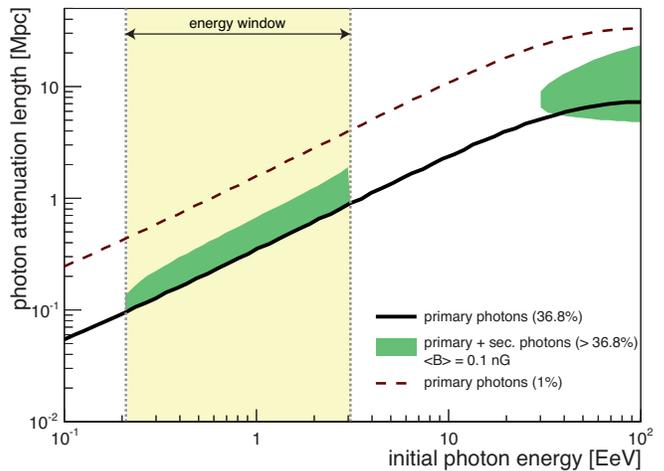}
\caption{Photon attenuation length as a function of the initial primary energy. The thick black line indicates the attenuation length (survival probability $e^{-1} \simeq 36.8\%$) and the dashed line indicates a reduced survival probability of 1\%. The energy range of this paper between $10^{17.3}$~eV and $10^{18.5}$~eV is indicated by the yellow shaded region and the vertical dotted lines. The expected increase of the observable distance by including secondary photons (detected in the energy range of this paper and with less than 1$^\circ$ deflection with respect to the primary photon) is shown as the green area using an average magnetic field strength of 0.1~nG.}
\label{fig.horizon}
\end{figure}

The detectable source distance is limited by interactions of UHE photons with low-energy background photons in pair or double-pair production processes. The attenuation length, i.e., the distance at which the survival probability has dropped to $e^{-1} \simeq 36.8\%$, depends on the energy of the UHE photon. The expected attenuation length for photons in interactions with the cosmic microwave background (dominating) and with radio \citep{Protheroe:1996si} and infrared \citep{Kneiske:2003tx} photon fields is shown as the solid black line in Figure \ref{fig.horizon}. In the energy range of the dataset, the attenuation length varies between 90~kpc~ at $10^{17.3}$~eV and 900~kpc at $10^{18.5}$~eV encompassing Galactic and nearby extragalactic sources. Requiring only a survival probability of 1\% the attenuation length increases, as indicated by the dashed line in Figure \ref{fig.horizon}, extending to a distance of a few Mpc at the highest energies considered and hence also including the nearest active galactic nucleus Centaurus A, which lies at less than 4~Mpc. It should be noted that the observable distance may increase further if including also secondary photons, i.e., UHE photons converted back from electrons via inverse Compton scattering within the electromagnetic cascading process taking place outside the Galaxy. However, to be useful for point source searches, these photons must be in the energy range and have parent electrons that have not been deflected more than 1$^\circ$ in ambient magnetic fields (see Section \ref{sec:Analysis}). The expected increase of the maximum observable distance in a Kolmogorov-type turbulent field with a mean magnetic field strength $\langle B \rangle = 0.1$~nG is shown as the green shaded area in Figure \ref{fig.horizon}, applying a three-dimensional CRPropa 3 simulation \citep{Batista:2016yrx}. Note that even primary photon energies outside the energy window above $10^{18.5}$~eV become visible if there is sufficient time to form the electromagnetic cascade. However, these results are very sensitive to the uncertain extragalactic magnetic field assumption, e.g., the maximum observable distance would drop to the values corresponding to the primary photon line if the mean magnetic field were $\langle B \rangle > 1$~nG, since in this case the electrons would be largely deflected.

Since there is a close connection between hadronic production processes for photons and neutrons, any candidate source of neutrons is also a candidate source of photons. As a consequence this analysis adopts the Galactic point source target sets defined in \cite{AugerTargetedNeutron} but adds the new H.E.S.S.\ unidentified sources reported in \cite{HESSDeilGP}. The Galactic source classes are millisecond pulsars (msec PSRs), $\gamma$-ray pulsars ($\gamma$-ray PSRs), low-mass and high-mass $X$-ray binaries (LMXBs and HMXBs), H.E.S.S.\ Pulsar Wind Nebulae (PWNe), other H.E.S.S.\ identified and unidentified sources, microquasars, magnetars, and the Galactic center. To retain independent target sets a candidate source that appears in two or more sets is kept only in the most exclusive set. Because the maximum observable distance of EeV photons is greater than that for EeV neutrons, two additional extragalactic target sets are included in this analysis.  One set consists of three powerful gamma-ray emitters in the Large Magellanic Cloud (LMC) at a distance of $\sim 50$~kpc \citep{HESSLMC}. The core region of Centaurus A (Cen A) is, by itself, the second extragalactic target set. The 12 source classes collectively include 364 individual candidate source directions.

\section{Analysis Method} 
\label{sec:Analysis}

To reduce the contamination of hadronic background events, only air showers similar to the photon expectation are selected using the multivariate method of Boosted Decision Trees~\citep{Breimanbook,Schapire1990} trained with Monte Carlo simulations of showers produced by photon and proton primaries. For a fixed primary energy, photon induced air showers have, on average, a delayed shower development and fewer muons (mostly electromagnetic component) compared to hadron-induced showers. As in the previous photon search paper, five different mass-sensitive observables are used: the depth of shower maximum $X_{\rm max}$ (from FD, being sensitive to delayed shower development), reduced $\chi^2$ and normalized energy of the Greisen fit to the longitudinal profile (from FD, being sensitive to the electromagnetic component), $S_b$-parameter~\citep{Ros2011140} (from SD, being sensitive to the slope of the lateral distribution of the shower, and hence to the muonic content), and the ratio of the early arriving to the late arriving integrated signal in the detector with the strongest signal (from SD, being sensitive to the muonic component and to the delayed shower development).

The optimized cut in the multivariate output distribution for a specific candidate source direction $i$ depends on the expected number of isotropic background events $b_i$. This number is calculated by applying the scrambling technique \citep{Scrambling1990}, and naturally takes into account detector efficiencies and aperture features by assigning arrival times and arrival directions, binned for each telescope, randomly from measured events. This procedure is repeated 5000 times and the mean number of arrival directions within a target is then used as the expected isotropic background count. For each target direction we use a top-hat counting region of 1$^\circ$. Details of this multivariate cut selection and counting procedure are given in \cite{BlindPhoton}. Averaging over all 364 target directions, the multivariate cut is expected to retain 81.4\% of photons while rejecting 95.2\% of background hadrons. After applying the cut, the total number of recorded events from all of the targets is reduced from 11,180 to 474. 

Each target set is tested with and without statistical weights. The weight $w_i$ is assigned to each target $i$ in a target set proportional to the measured electromagnetic flux $f_i$ in the catalog and proportional to the directional photon exposure $\epsilon_i$ of the Pierre Auger Observatory based on \cite{ExposureAuger}. Relative attenuation differences from candidate sources of the same class can be neglected given an interaction length between 90 and 900~kpc of primary photons in the energy range considered (see Figure \ref{fig.horizon}). The sum of weights in each set is normalized to 1 (see \cite{AugerTargetedNeutron}): 
\begin{equation}
w_i = \frac{f_i \cdot \epsilon_i}{\sum_i f_i \cdot \epsilon_i}~.
\end{equation}

A $p$-value $p_i$ is assigned to each candidate source of a target set as follows. The $p$-value for the target $i$ is defined by $p_i \equiv [{\rm Poisson} (n_i, b_i) + {\rm Poisson} (n_i + 1, b_i)]/2$, where Poisson$(n_i, b_i)$ is the probability of getting $n_i$ or more arrival directions in the target when the observed value is $n_i$, and the expected number from the background is $b_i$. Averaging the values for $n$ and $n + 1$ avoids a bias toward low or high $p$-values for pure background fluctuations. 

The combined weighted probability $\mathcal{P}_{w}$ is the fraction of isotropic simulations yielding a weighted product $\prod_{i} p_{i, \rm iso}^{w_i}$ that is not greater than the measured weighted product $\prod_i p_i^{w_i}$:
\begin{equation}
\mathcal{P}_{w} = {\rm Prob} \left( \prod_{i} p_{i, \rm iso}^{w_i} \leq \prod_i p_i^{w_i} \right)~,
\end{equation}
where $p_{i, \rm iso}$ denotes the $p$-value of target $i$ in an isotropic simulation. The combined unweighted probability $\mathcal{P}$ is given by the same formula with $w_i = 1$ for all targets (see \cite{AugerTargetedNeutron}).

\section{Results} 
\label{sec:Results}

The results for the combined analysis for each of the 12 target sets are shown in Table \ref{tab:Targets}, along with detailed information about the target that has the smallest $p$-value in each set. In addition to the direction of the candidate source, the measured and expected numbers of events within an opening angle of 1$^\circ$ are given along with the required number of events for a $3\sigma$ observation. In the last two columns are the minimum $p$-value of the target set ($p$) and the penalized $p$-value $p^* = 1- (1-p)^N$, which is the chance probability that one or more of the $N$ candidate sources in the target set would have a $p$-value less than $p$ if the $N$ $p$-values were randomly sampled from the uniform probability distribution.

No combined $p$-value ($\mathcal{P}$ or $\mathcal{P}_{w}$) nor any individual target $p$-value has a statistical significance as great as $3\sigma$. Upper limits are therefore derived for the flux from the target of smallest $p$-value in each target set assuming an $E^{-2}$ photon spectrum and they are indicated in Table \ref{tab:Targets}. Upper limits on the photon flux from a point source $i$ are calculated as $f^{\rm 95\%}_i = n_i^{\rm Zech}/(n_{\rm inc} \cdot \epsilon_{i})$, where $n_i^{\rm Zech}$ is the upper limit, at the 95\% confidence level, on the number of photons using Zech's definition \citep{ZECH1989608}, $n_{\rm inc}=0.9$ is the expected signal fraction within the search window, and $\epsilon_i$ is the directional photon exposure.

Various sources of systematic uncertainties have been investigated in \cite{BlindPhoton}. The main contribution arises from the unknown photon spectral index due to the associated change in the directional photon exposure. Differences in the particle flux upper limit of $-34\%$ and $+51\%$ have been estimated when changing the photon spectral index from 2.0 to 1.5 or 2.5, respectively. Considering the background rejection, differences in the hadronic interaction models change the particle flux upper limits by, on average, -9\% when using EPOS-LHC \citep{Pierog:2013ria} for proton simulations instead of QGSJET-01c \citep{Kalmykov:1989br}.

\begin{figure}[t!]
\centering
\includegraphics[width=8.6cm]{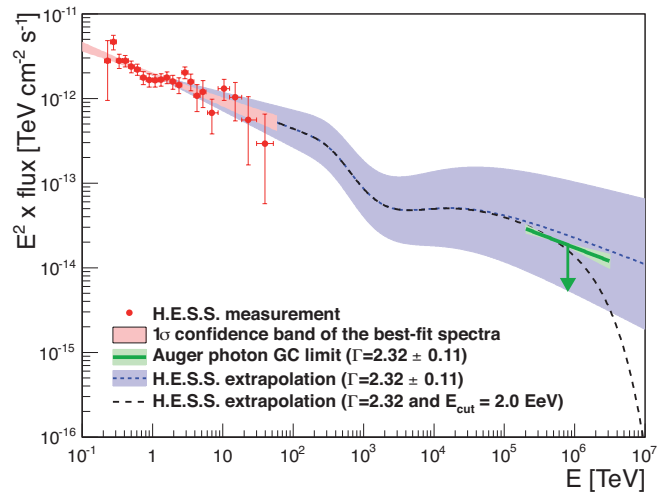}
\caption{Photon flux as a function of energy from the Galactic center region. Measured data by H.E.S.S.\ are indicated, as well as the extrapolated photon flux at Earth in the EeV range, given the quoted spectral indices (\cite{HESSPevatron}; conservatively the extrapolation does not take into account the increase of the $p$--$p$ cross-section toward higher energies). The Auger limit is indicated by a green line. A variation of the assumed spectral index by $\pm 0.11$ according to systematics of the H.E.S.S.\ measurement is denoted by the light green and blue band. A spectral index with cutoff energy $E_{\rm cut} = 2.0 \cdot 10^{6}$ TeV is indicated as well.}
\label{fig.spectrum}
\end{figure}

In the following, the limit on the Galactic center is examined in more detail. This is of particular interest, as the H.E.S.S.\ collaboration recently reported an indication of the acceleration of PeV protons from this region \citep{HESSPevatron}. H.E.S.S.\ measured the diffuse $\gamma$-ray emission following a power-law spectrum according to ${\rm d}N / {\rm d} E = \Phi_1 E^{-\Gamma}$ with a spectral index of $\Gamma = 2.32 \pm 0.05_{\rm stat} \pm 0.11_{\rm syst}$ and flux normalization $\Phi_1 = (1.92 \pm 0.08_{\rm stat} \pm 0.28_{\rm syst}) \times 10^{-12}$~TeV$^{-1}$~cm$^{-2}$~s$^{-1}$ without a cutoff or break up to tens of TeV \citep{HESSPevatron}. Since the results in Table \ref{tab:Targets} are based on a photon spectral index of $\Gamma = 2$ the limit is recalculated assuming  $\Gamma = 2.32$ resulting in a particle flux upper limit at the 95\% confidence level of the Galactic center region of $J_{\rm int}^{\rm UL} = 0.034$~km$^{-2}$~yr$^{-1}$.  As can be seen in Figure \ref{fig.spectrum}, the current photon flux upper limit can severely constrain the allowed parameter space for a flux continuation to EeV energies. This extrapolation takes into account interactions with the cosmic microwave background (dominating) and with radio \citep{Protheroe:1996si} and infrared \citep{Gilmore2012} photon fields. Furthermore, assuming a power law with an exponential cutoff of the form ${\rm d}N / {\rm d} E = \Phi_1 E^{-\Gamma} \times \exp{(-E / E_{\rm cut})}$ an upper limit of the cutoff energy $E_{\rm cut} = 2.0$~EeV can be placed by setting $\int_{E_1}^{E_2} \Phi_1 E^{-\Gamma} \times \exp{(-E / E_{\rm cut})~{\rm d}E} = J_{\rm int}^{\rm UL}$ with $E_1 = 10^{17.3}$~eV and $E_2 = 10^{18.5}$~eV and solving for $E_{\rm cut}$. The corresponding cutoff spectrum is also given by the dashed line in Figure \ref{fig.spectrum}.

\section{Discussion} 
\label{sec:Summary}

No target class reveals compelling evidence for photon-emitting sources in the EeV energy regime. For the 12 sets, the minimum combined weighted probability $\mathcal{P}_w$ is 0.12.  With 12 trials, one expects a $\mathcal{P}_w$-value at least that small to occur by chance, with 78\% confidence. The minimum unweighted $\mathcal{P}$-value, 0.14, is similarly not statistically significant. There is also no evidence for one outstanding target in any target set. The minimum penalized $p$-value $p^*$ in the 12 sets is 0.221. The null result holds also against the hypothesis that only a subset of some target class contributes a photon flux. This has been tested by calculating combined $\mathcal{P}$-values scanning only over the most significant, i.e., the smallest $p$-value, targets in the catalog. 

The results presented in this paper complement previous results published by the Pierre Auger Collaboration searching for neutrons at higher energies using SD data \citep{NeutronBlind0004-637X-760-2-148,AugerTargetedNeutron}, and photons using hybrid data \citep{BlindPhoton}, by restricting the photon search to 12 predefined target classes. Flux upper limits from photon point sources constrain the continuation of measured TeV fluxes to EeV energies, as shown for the particular case of the Galactic center introducing an upper limit of the cutoff photon energy of $E_{\rm cut} = 2.0$~EeV.

The discovery of photon fluxes from any target set or
individual targets in this study would have proved that EeV protons
are being accelerated at discrete sources within the Galaxy or its
neighborhood. The null results reported here leave open the possibility that EeV
protons, as observed on Earth, are of extragalactic origin. Some
support for that hypothesis was noted in the large-scale anisotropy
analysis of Auger data~\citep{Auger2041-8205-762-1-L13}. It is important to note, however, that
the absence of detectable photon fluxes, as reported here, does not
exclude the production of EeV protons within the Galaxy. The derived flux limits are time-averaged values. EeV photons might be produced in transient sources, such as gamma-ray bursts or supernovae, or aligned in jets not pointing to us. An alternative explanation is that EeV protons escape from a source more freely than protons that produce TeV photon fluxes, and the production of EeV photons is thereby too meager to be detectable in the present study.

With the detector upgrade AugerPrime \citep{Engel:2015ibd,Aab:2016vlz} the photon-hadron separation will be further improved, allowing an increased sensitivity for photon point sources.

% created on 2016-12-02

\section*{Acknowledgments}

\begin{sloppypar}
The successful installation, commissioning, and operation of the Pierre Auger Observatory would not have been possible without the strong commitment and effort from the technical and administrative staff in Malarg\"ue. We are very grateful to the following agencies and organizations for financial support:
\end{sloppypar}

\begin{sloppypar}
Argentina -- Comisi\'on Nacional de Energ\'\i{}a At\'omica; Agencia Nacional de Promoci\'on Cient\'\i{}fica y Tecnol\'ogica (ANPCyT); Consejo Nacional de Investigaciones Cient\'\i{}ficas y T\'ecnicas (CONICET); Gobierno de la Provincia de Mendoza; Municipalidad de Malarg\"ue; NDM Holdings and Valle Las Le\~nas, in gratitude for their continuing cooperation over land access; Australia -- the Australian Research Council; Brazil -- Conselho Nacional de Desenvolvimento Cient\'\i{}fico e Tecnol\'ogico (CNPq); Financiadora de Estudos e Projetos (FINEP); Funda\c{c}\~ao de Amparo \`a Pesquisa do Estado de Rio de Janeiro (FAPERJ); S\~ao Paulo Research Foundation (FAPESP) Grants No.\ 2010/07359-6 and No.\ 1999/05404-3; Minist\'erio de Ci\^encia e Tecnologia (MCT); Czech Republic -- Grant No.\ MSMT CR LG15014, LO1305, and LM2015038 and the Czech Science Foundation Grant No.\ 14-17501S; France -- Centre de Calcul IN2P3/CNRS; Centre National de la Recherche Scientifique (CNRS); Conseil R\'egional Ile-de-France; D\'epartement Physique Nucl\'eaire et Corpusculaire (PNC-IN2P3/CNRS); D\'epartement Sciences de l'Univers (SDU-INSU/CNRS); Institut Lagrange de Paris (ILP) Grant No.\ LABEX ANR-10-LABX-63 within the Investissements d'Avenir Programme Grant No.\ ANR-11-IDEX-0004-02; Germany -- Bundesministerium f\"ur Bildung und Forschung (BMBF); Deutsche Forschungsgemeinschaft (DFG); Finanzministerium Baden-W\"urttemberg; Helmholtz Alliance for Astroparticle Physics (HAP); Helmholtz-Gemeinschaft Deutscher Forschungszentren (HGF); Ministerium f\"ur Innovation, Wissenschaft und Forschung des Landes Nordrhein-Westfalen; Ministerium f\"ur Wissenschaft, Forschung und Kunst des Landes Baden-W\"urttemberg; Italy -- Istituto Nazionale di Fisica Nucleare (INFN); Istituto Nazionale di Astrofisica (INAF); Ministero dell'Istruzione, dell'Universit\'a e della Ricerca (MIUR); CETEMPS Center of Excellence; Ministero degli Affari Esteri (MAE); Mexico -- Consejo Nacional de Ciencia y Tecnolog\'\i{}a (CONACYT) No.\ 167733; Universidad Nacional Aut\'onoma de M\'exico (UNAM); PAPIIT DGAPA-UNAM; The Netherlands -- Ministerie van Onderwijs, Cultuur en Wetenschap; Nederlandse Organisatie voor Wetenschappelijk Onderzoek (NWO); Stichting voor Fundamenteel Onderzoek der Materie (FOM); Poland -- National Centre for Research and Development, Grants No.\ ERA-NET-ASPERA/01/11 and No.\ ERA-NET-ASPERA/02/11; National Science Centre, Grants No.\ 2013/08/M/ST9/00322, No.\ 2013/08/M/ST9/00728, and No.\ HARMONIA 5 -- 2013/10/M/ST9/00062; Portugal -- Portuguese national funds and FEDER funds within Programa Operacional Factores de Competitividade through Funda\c{c}\~ao para a Ci\^encia e a Tecnologia (COMPETE); Romania -- Romanian Authority for Scientific Research ANCS; CNDI-UEFISCDI partnership projects Grants No.\ 20/2012 and No.194/2012 and PN 16 42 01 02; Slovenia -- Slovenian Research Agency; Spain -- Comunidad de Madrid; Fondo Europeo de Desarrollo Regional (FEDER) funds; Ministerio de Econom\'\i{}a y Competitividad; Xunta de Galicia; European Community 7th Framework Program Grant No.\ FP7-PEOPLE-2012-IEF-328826; USA -- Department of Energy, Contracts No.\ DE-AC02-07CH11359, No.\ DE-FR02-04ER41300, No.\ DE-FG02-99ER41107 and No.\ DE-SC0011689; National Science Foundation, Grant No.\ 0450696; The Grainger Foundation; Marie Curie-IRSES/EPLANET; European Particle Physics Latin American Network; European Union 7th Framework Program, Grant No.\ PIRSES-2009-GA-246806; and UNESCO.
\end{sloppypar}

%\bibliography{biblio}{}
%\bibliographystyle{aasjournal}

\end{document}